\RequirePackage{fix-cm}
\documentclass[smallcondensed]{svjour3}     
\smartqed  
\usepackage{graphicx}
\usepackage{amsmath}
\usepackage{graphicx}
\usepackage{latexsym}
\usepackage{hyperref}
\newcommand{\ba}{\begin{eqnarray}}
\newcommand{\ea}{\end{eqnarray}}
\newcommand{\beq}{\begin{eqnarray}}
\newcommand{\eeq}{\end{eqnarray}}

\newcommand{\nn}{\nonumber}

\newcommand{\bs}{\boldsymbol} 
\newcommand{\es}{&=&}

\newcommand{\ts}{&\times&}
\begin{document}
\title{Three Dimensional Imaging of the Nucleon
}

\author{Jai More\and 
	Asmita Mukherjee \and 
	Sreeraj Nair
          }
  \institute{Jai More\at
      Indian Institute of Technology, Bombay, Mumbai-400076, India. \\
\email jaimore@iitb.ac.in\\
\\
       Asmita Mukherjee\at
      Indian Institute of Technology, Bombay, Mumbai-400076, India. \\
      \email asmita@phy.iitb.ac.in         \\
              \\     Sreeraj Nair \at Indian Institute of Science Education and Research, Bhopal, MP-462030, India.\\
                   \email sreeraj\_nair@iiserb.ac.in
}
%
%
\date{Received: date / Accepted: date}
\maketitle
\begin{abstract}
We study the Wigner distributions of quarks and gluons in light-front dressed quark model using the overlap of light front wave 
functions (LFWFs). We take the target to be a dressed quark,  this is  a composite spin$-1/2$ state of quark dressed with a gluon. 
This state allows us to calculate the quark and gluon Wigner distributions analytically in
terms of LFWFs using Hamiltonian perturbation theory. 
We analyze numerically the Wigner distributions of quark and gluon and report their nature in the contour plots. 
We use an improved numerical technique to remove the cutoff dependence of the Fourier transformed integral over ${\bs \Delta}_\perp$. 
\end{abstract}
\section{Introduction}\label{intro}
The most demanding issue in QCD is to have the best possible theoretical description of hadrons in terms of quarks and gluons.
One of the possible approaches would be to study hadrons in terms of so-called generalized parton correlation functions (GPCFs)\cite{Meissner09}. 
The GPCFs depend on the four-momentum of the partons and the momentum transfer to the hadrons and thus are rich in information about
the partonic structure of the nucleon. Significantly, GPCFs are interesting as they are connected to Wigner distributions \cite{Wigner32}  via 
Fourier transform. Although the Wigner distributions are quasi-probabilistic, they reduce to probability distributions on integration
over transverse position and/or transverse momentum space coordinates.
Thus by taking the appropriate limit, one can obtain impact parameter dependent distributions (IPDs) and transverse momentum 
dependent distributions (TMDs) from the Wigner distributions. Wigner distributions are becoming prevailing
and recently there has been an intensive study that proposed their experimental measurements \cite{Bhattacharya17,Hatta16,HattaXiao16,Zhou16,Hagiwara16}.

We study the quark and gluon Wigner distributions in a dressed quark model using Hamiltonian perturbation theory.
Here the target state is a single quark dressed with a gluon; this contains gluonic degrees of freedom
within a perturbative model. The state is expanded in terms of the multiparton wavefunctions 
called light front wavefunctions (LFWFs). 

The manuscript is arranged in the following manner. In Sec. \ref{sec:1} we discuss quark and gluon Wigner distributions 
and definitions. The complete discussion for different polarization configuration of the Wigner distributions in dressed 
quark model can be looked in Ref \cite{More17,More18}.  In our model, we obtain 8 independent quark Wigner distributions and 
6 independent gluon Wigner distributions.
Here we limit ourselves to discuss only the unpolarized quarks and gluons for different polarization of the target state at leading twist.
Then in Sec. \ref{sec:2} we discuss numerical results and plots for some of the independent Wigner distributions of quarks and gluons.
Finally, our conclusion is given in Sec. \ref{conclusion}.
\section{Wigner distributions of quarks and gluons}
\label{sec:1}

The Wigner distribution of quarks and gluons can be defined as the following Fourier transform
\cite{Meissner09,Lorce11}
\ba
W^{q/g}_{s\,s'}({\bs b}_{\perp},{\bs k}_{\perp},x) = \int \frac{d^2 \Delta_{\perp}}{(2\pi)^2} e^{-i {\bs \Delta}_{\perp}.{\bs b}_{\perp}}
\int \!\!\frac{dz^{-}d^{2} {\bs z}_{\perp}}{2(2\pi)^3}e^{i k.z}
 \Big{\langle}p',s' \Big{|}
\mathcal{O}^{q/g}(z) \Big{|}
p,s\Big{\rangle }  \Big{|}_{z^{+}=0}
\ea
where ${\bs b_\perp}$ is the impact parameter conjugate to ${\bs \Delta}_\perp$, which is momentum transfer 
in the transverse direction. We take the momentum transfer in the
longitudinal direction to be zero. Here the superscript $q(g)$ indicates the quark (gluon).
The corresponding operators are defined as follows:
\ba
\mathcal{O}^{q}(z) \es 
\overline{\psi}(-\frac{z}{2})\Omega \Gamma \psi(\frac{z}{2}) \\
\mathcal{O}^{g} (z)\es 
\frac{1}{xp^+}\Gamma^{ij} F^{+i}\Big( -\frac{z}{2}\Big) F^{+j}\Big( \frac{z}{2}\Big),
 \ea
where $\Omega $ is the gauge link which is set to unity and we use the light cone gauge $A^+=0$.
The twist-two Dirac operators are $\Gamma=\{\gamma^+, \gamma^+\gamma^5,  i \sigma^{+1}\gamma^{5}, 
i \sigma^{+2}\gamma^{5}\}$ and they correspond to  Wigner distributions corresponding to an unpolarized, 
longitudinally polarized and transversely polarized dressed quark respectively.  
For the gluon case the operator at twist two are \cite{Lorce13}
 $\Gamma^{ij}=\{\delta_\perp^{ij},-i\epsilon^{ij}_{\perp},\Gamma^{RR},\Gamma^{LL}\}$ where $L(R)$ are left(right) polarization of the gluon. 
  The target dressed quark state can be written as a series in Fock space. The dynamics of each state in this Fock space is captured by the $n$-particle LFWFs.
\ba
  \Big{| }p^{+}, {\bs p}_{\perp}, s \Big{\rangle} &=& \Phi^{s}(p) b^{\dagger}_{s}(p) | 0 \rangle +
 \sum_{s_1 s_2} \int \frac{dp_1^{+}d^{2}{\bs p}_1^{\perp}}{ \sqrt{16 \pi^3 p_1^{+}}}
 \int \frac{dp_2^{+}d^{2}{\bs p}_2^{\perp}}{ \sqrt{16 \pi^3 p_2^{+}}} \sqrt{16 \pi^3 p^{+}}
  \nn \\[1.5ex] 
 &\times&\delta^3(p-p_1-p_2)\Phi^{s}_{s_1 s_2}(p;p_1,p_2) b^{\dagger}_{s_1}(p_1) 
 a^{\dagger}_{s_2}(p_2)  | 0 \rangle ,
 \ea
$\Phi^{s}(p)$ gives the wave function normalization; $\Phi^{s}_{s_1 s_2}(p;p_1,p_2)$ is the quark gluon state LFWF. $\Phi^{s}_{s_1 s_2}(p;p_1,p_2)$ 
gives the probability amplitude to find a bare quark (gluon) with momentum $p_1 (p_2)$ and helicity $s_1 (s_2)$ inside the dressed quark.
Using the Jacobi momenta:
$\,\, k_{i}^{+} = x_{i}P^+ ~\text{and}~~~ {\bs k}_{i}^{\perp} ={\bs q}_{i}^{\perp} +  x_{i}{\bs P}^{\perp} $\\
so that 
\ba
\sum_i x_i=1,~~~~~\sum_i {\bs q}_{i\perp}=0;
\ea 
the two particle LFWF can be written in terms of boost invariant LFWF as
\ba
\sqrt{P^+}\Phi(p; p_1, p_2) = \Psi(x_{i},{\bs q}_{i}^{\perp}).
\ea
The two particle LFWF is given by \cite{Hari99} 
\ba\label{2particlelfwf}
\Psi^{sa}_{s_1 s_2}(x,{\bs q}^{\perp})&=& 
\frac{1}{\Big[m^2 - \frac{m^2 + ({\bs q}^{\perp})^2 }{x} - \frac{({\bs q}^{\perp})^2}{1-x} \Big]}
\frac{g}{\sqrt{2(2\pi)^3}} T^a \chi^{\dagger}_{s_1} \frac{1}{\sqrt{1-x}}\nn\\
&\times&
 \Big[ -\frac{2{\bs q}^{\perp}}{1-x}   -  \frac{({\bs \sigma}^{\perp}.{\bs q}^{\perp}){\bs \sigma}^{\perp}}{x}
+\frac{im~{\bs \sigma}^{\perp}(1-x)}{x}\Big]
\chi_s ({\bs \epsilon}^{\perp}_{s_2})^{*}.
\ea

For the quark operator with Dirac bilinear as $\gamma^+$ we can obtain quark GTMDs in a dressed quark state in terms of two particle LFWFs as follows;
\ba\label{U}
W_{s\, s'}^{q} ({\bs \Delta}_{\perp},{\bs k}_{\perp},x) & =&\sum_{\lambda_1',\lambda_{1},\lambda_2} 
\Psi^{*s'}_{\lambda_{1}' \lambda_2}(x,{\bs q'}^{\perp}) \chi_{\lambda_1'}^{\dagger} \chi_{\lambda_1}
\Psi^{s}_{\lambda_1 \lambda_2}(x,{\bs q}^{\perp}),
\ea
where $x$ is the longitudinal momentum fraction carried by the quark and ${\bs q}^\perp,\, {\bs q}^{'\perp}$ denotes the transverse momentum of quark that can be expressed in the symmetric frame \cite{More18}. 

Using Eq.~\ref{2particlelfwf} and taking the Fourier transform of GTMDs with
respect to $\Delta_\perp$ in Eq. \ref{U}, we procure the analytical expressions for the Wigner distributions of an unpolarized quark inside a target state with different polarizations:
\ba
W_{UU}^q(x,{\bs k}_\perp,{\bs b}_\perp)\es N\int\frac{d^2\Delta_{\perp}}{2(2\pi)^2}\frac{\cos({\bs \Delta}_{\perp}\cdot{\bs b}_{\perp})}{D({\bs q}_\perp)D({\bs q'}_\perp)}
\nn\\
\ts\Big[\frac{\Big(4k_\perp^2-\Delta_\perp^2 (1-x)^2\Big)(1+x^2)}{x^2(1-x)^3}+\frac{4 m^2(1-x)}{x^2}\Big]\label{rhouu} \\
W_{LU}^q(x,{\bs k}_\perp,{\bs b}_\perp)\es N \!\!\!\int\!\!\! \frac{d^2\Delta_{\perp}}{2(2\pi)^2}\! \frac{\sin({\bs \Delta}_{\perp}\cdot {\bs b}_{\perp})}{D({\bs q}_\perp)D({\bs q'}_\perp)}
 \Big[\frac{4\Big(k_y\Delta_x-k_x\Delta_y\Big)(1+x)}{x^2(1-x)}\Big]\label{rhoul}
\\
W_{TU}^q(x,{\bs k}_\perp,{\bs b}_\perp)\es N \int \frac{d^2\Delta_{\perp}}{2(2\pi)^2}\, \frac{\sin({\bs \Delta}_{\perp}\cdot{\bs b}_{\perp})}{D({\bs q}_\perp)D({\bs q'}_\perp)}
 \Big[\frac{4m \Delta_x}{x}\Big],\label{rhotu}
\ea
where $W^q_{UU},W^q_{LU}$ and $W^q_{TU} $ are the Wigner distributions for an unpolarized quark in an unpolarized, 
longitudinally polarized and transversely polarized target state respectively. Here we chose the transverse polarization to be in the $x$ direction.

Similarly, for the operator $\delta_\perp^{ij}$ we obtain gluon GTMDs in terms of overlap of LFWFs as
\ba\label{W1}
\mathcal{W}^{g}_{s\, s'}(x,{\bs k _\perp},{\bs \Delta _\perp})=-\!\!\!\sum_{\sigma_1, \lambda_{1},\lambda_{2}}
\!\!\!\Psi^{*s'}_{\sigma_1 \lambda_{1} }(\hat{x},\hat{{\bs q}}'_{\perp}) \Psi^{s}_{\sigma_1 \lambda_2}(\hat{x},\hat{{\bs q}}_{\perp})
\Big(\epsilon_{\lambda_2}^{1} \epsilon_{\lambda_1}^{*1}+\epsilon_{\lambda_2}^{2} \epsilon_{\lambda_1}^{*2}\Big),
\ea
where $\hat{x}$ is the longitudinal momentum fraction carried by the quark and $\hat{\bs q }^\perp,\, \hat{\bs q}'^\perp$ denotes the transverse momentum of quark that can be expressed in the symmetric frame. 
The gluon longitudinal momentum fraction and transverse momentum are related to them in the following way:
$\hat{x}=(1-x)$ and $\hat{\bs q}_\perp=-{\bs q_\perp}$.

In a similar manner we obtain Wigner distribution of unpolarized gluon inside the dressed quark state with different polarizations as $W^g_{UU},W^g_{LU}$ and $W^g_{TU} $. Their analytical expressions are follows:
\ba
W_{UU}^g(x, {\bs k _\perp}, {\bs b _\perp})&=&-N\int \frac{d^2 {\bs \Delta _\perp}}{2(2\pi)^2} \frac{\cos({\bs \Delta_\perp}\cdot{\bs b_\perp})}{D({\bs q_\perp})D({\bs q'_\perp})}\nn\\
&\times&\left[\frac{4 m^2 x^4+(x^2-2x+2) \Big(4{\bs k _\perp}^2-{\bs \Delta _\perp}^2 (1-x)^2\Big)}{(1-x)^2 x^3}\right]\label{uu}\\[2ex]
W_{LU}^g(x, {\bs k_\perp}, {\bs b_\perp})&=&N \!\!\int\!\! \frac{d^2 {\bs \Delta _\perp}}{2(2\pi)^2} \frac{\sin({\bs \Delta_\perp}\cdot{\bs b_\perp})}{D({\bs q_\perp})D({\bs q'_\perp})}
\left[\frac{4 (2-x) (\Delta_y k_x-\Delta_x k_y)}{(1-x) x^2}\right]\label{lu}\\[2ex]
W_{T U}^g(x, {\bs k _\perp}, {\bs b _\perp})&=&N \int \frac{d^2 {\bs \Delta _\perp}}{2(2\pi)^2} \frac{\sin({\bs \Delta_\perp}\cdot{\bs b_\perp})}{D({\bs q_\perp})D({\bs q'_\perp})}
\left[\frac{4 m \Delta_x}{x}\right]\label{tu},
\ea
where $W^g_{UU},W^g_{LU}$ and $W^g_{TU} $ are the Wigner distribution for an unpolarized gluon in an unpolarized, longitudinally polarized and transversely polarized target state respectively. Here also we chose the transverse polarization of target state to be in the $x$ direction. 
It should be noted that distribution of unpolarized quark and gluon inside a transversely polarized target looks identical as seen from Eqs.~\ref{rhotu} and \ref{tu}. However, in Eq.~\ref{rhotu}, $x$ is longitudinal momentum fraction of quark while  
in Eq~\ref{tu}, $x$ is the longitudinal momentum fraction of gluon. The explicite expressions for  $D(q_\perp)$ and  $D(q'_\perp)$ 
 in Eqs. (\ref{rhouu})-(\ref{rhotu}) and (\ref{uu})-(\ref{tu})  can be found in reference Ref \cite{More17,More18}. 

\section{Numerical analysis and plots}
The five dimensional Wigner distributions are studied in the transverse phase space (${\bs k _\perp}, {\bs b _\perp}$) by integrating out the longitudinal momentum fraction  ($x$).
\label{sec:2}\begin{figure}[h!]
 \centering 
(a)\includegraphics[width=5cm,height=3.5cm]{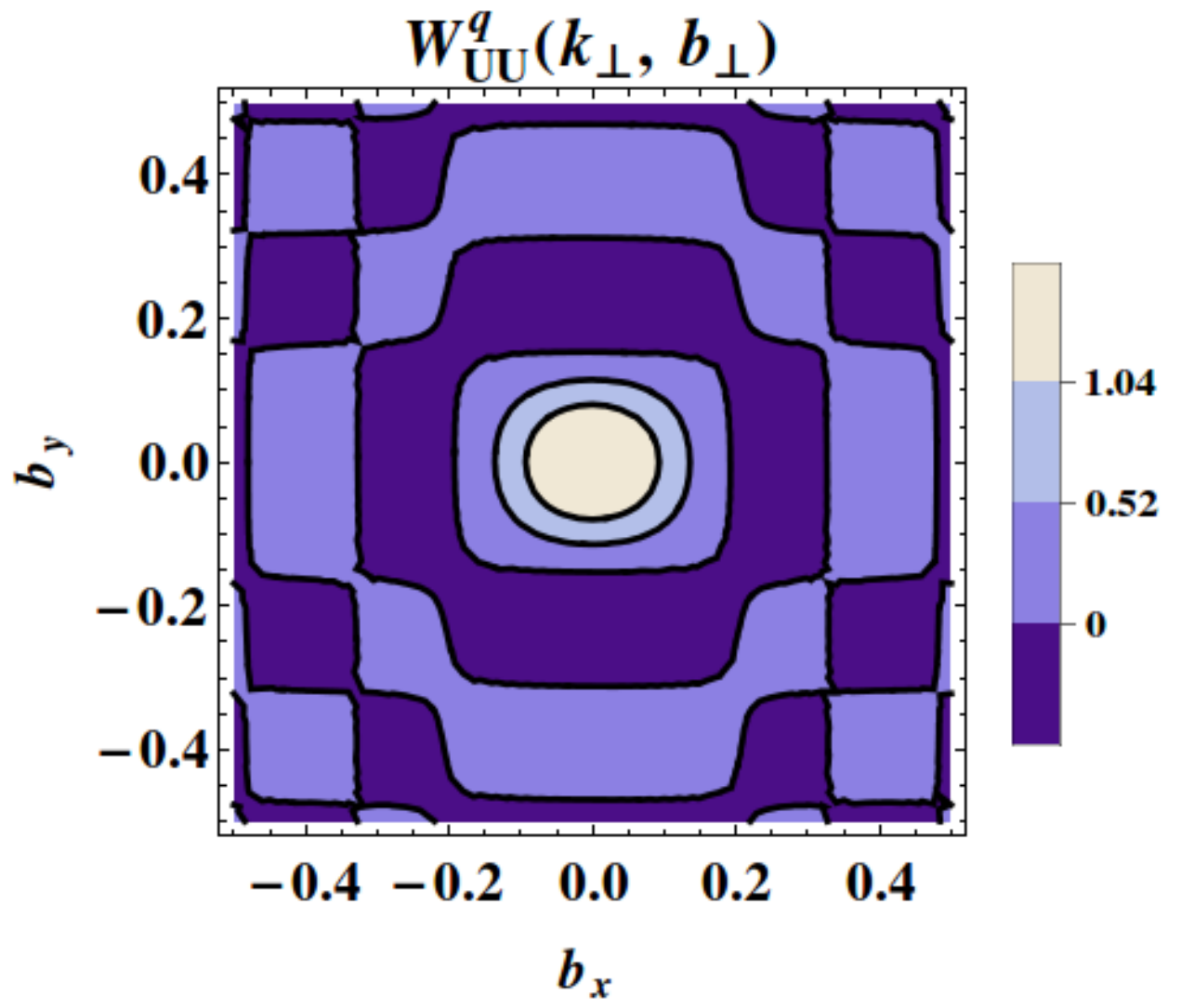} 
(d)\includegraphics[width=5cm,height=3.5cm]{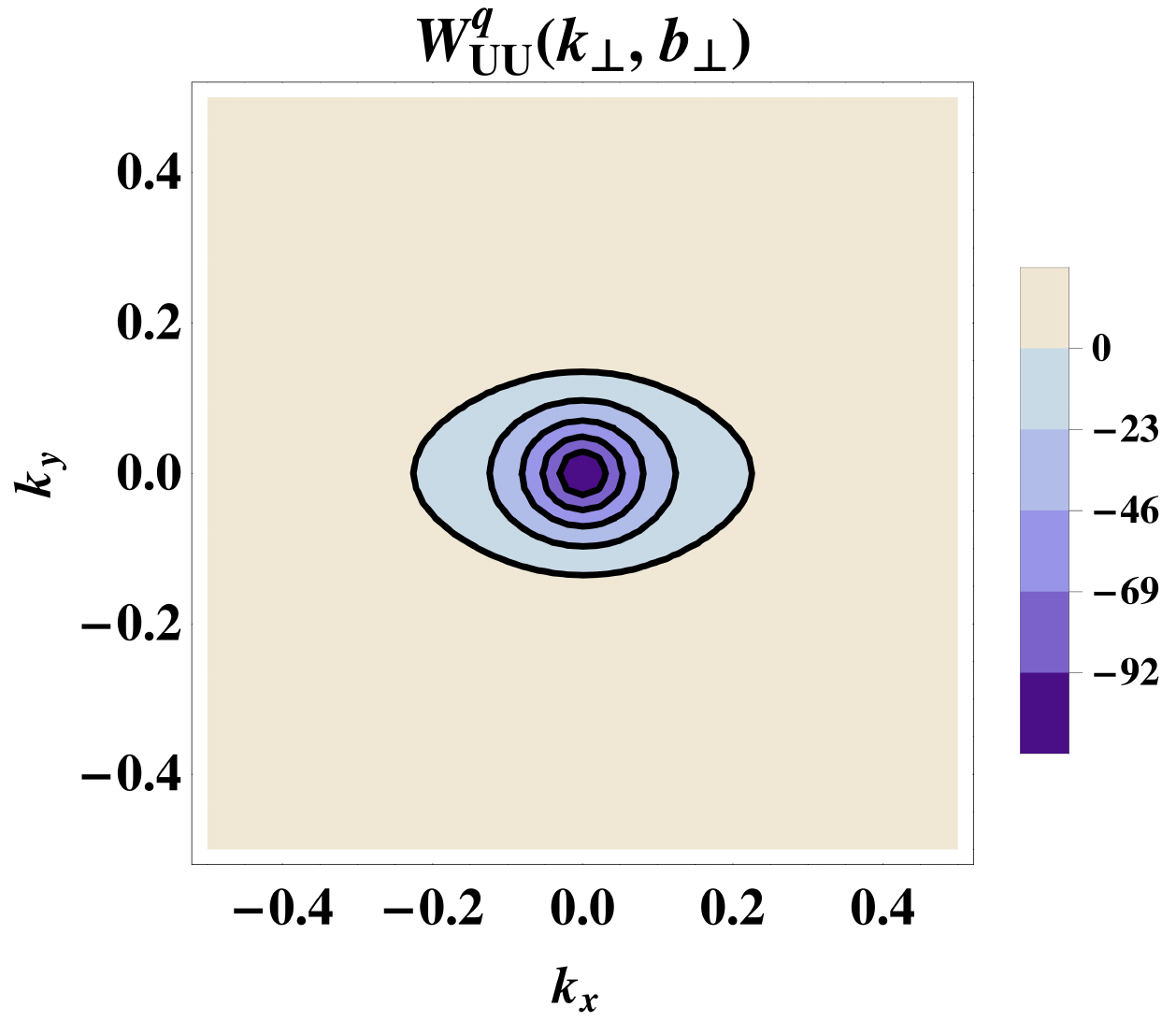}\\[1ex]
(b)\includegraphics[width=5cm,height=3.5cm]{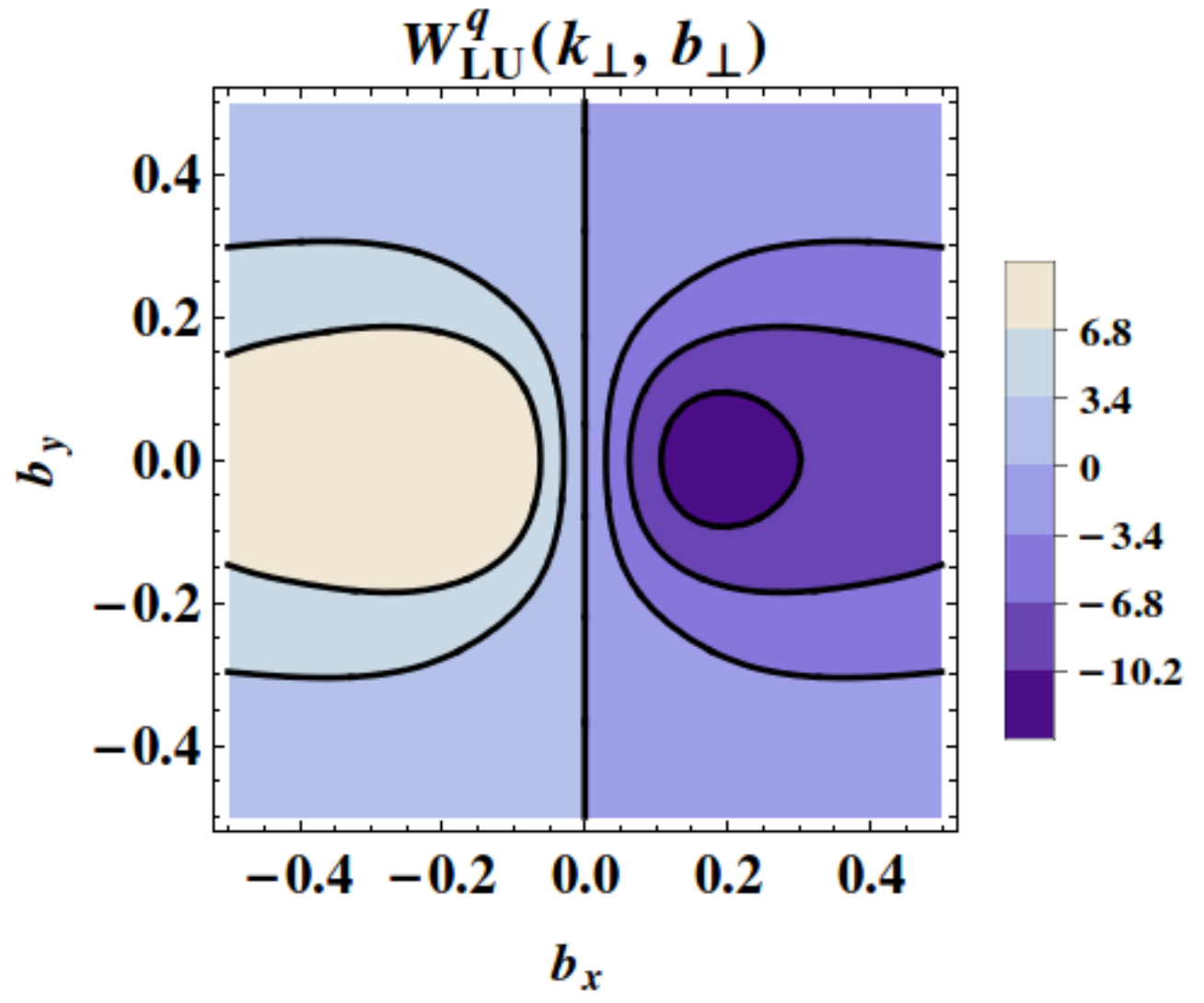}
(e)\includegraphics[width=5cm,height=3.5cm]{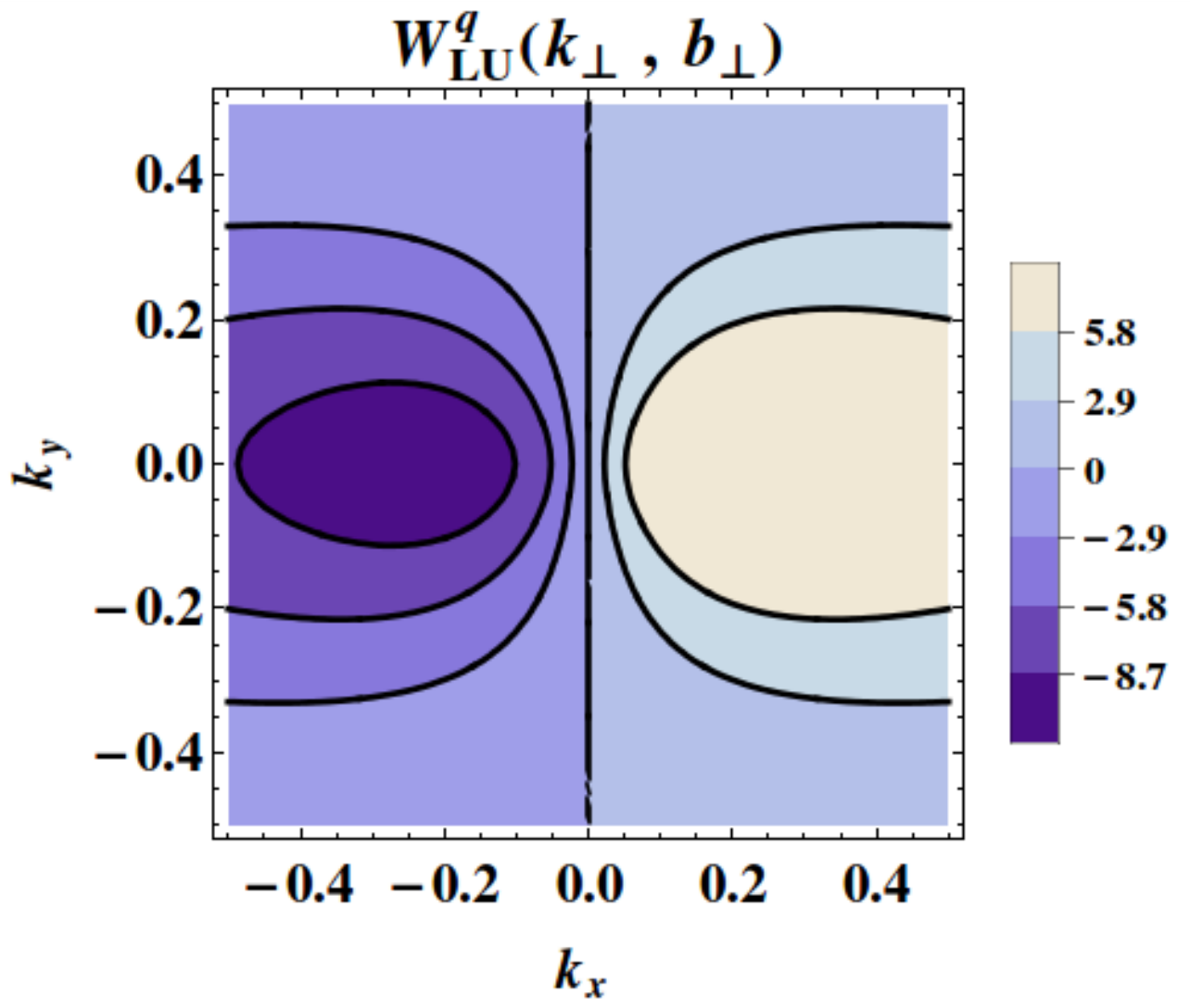}\\[1ex]
(c)\includegraphics[width=5cm,height=3.5cm]{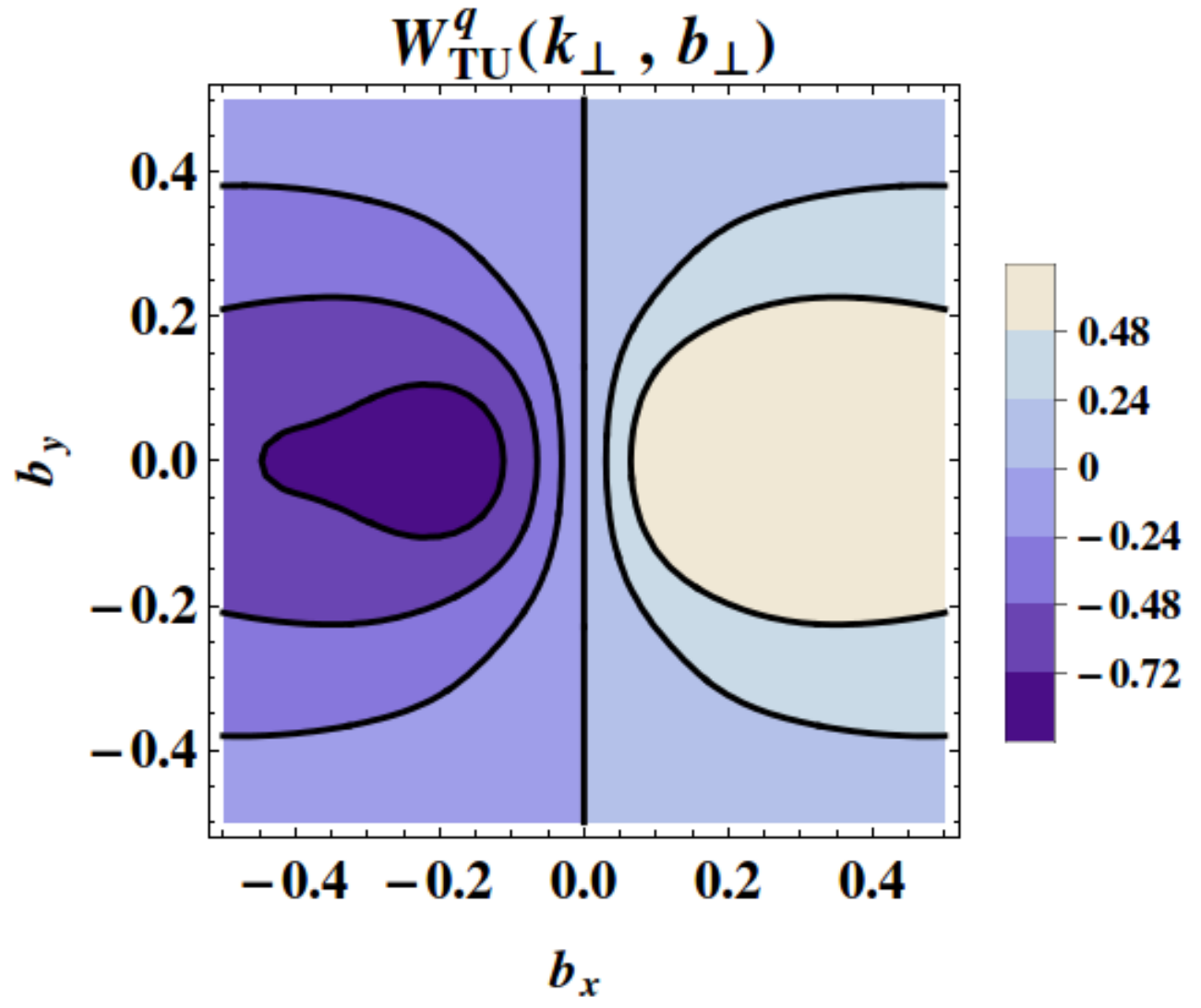}
(f)\includegraphics[width=5cm,height=3.5cm]{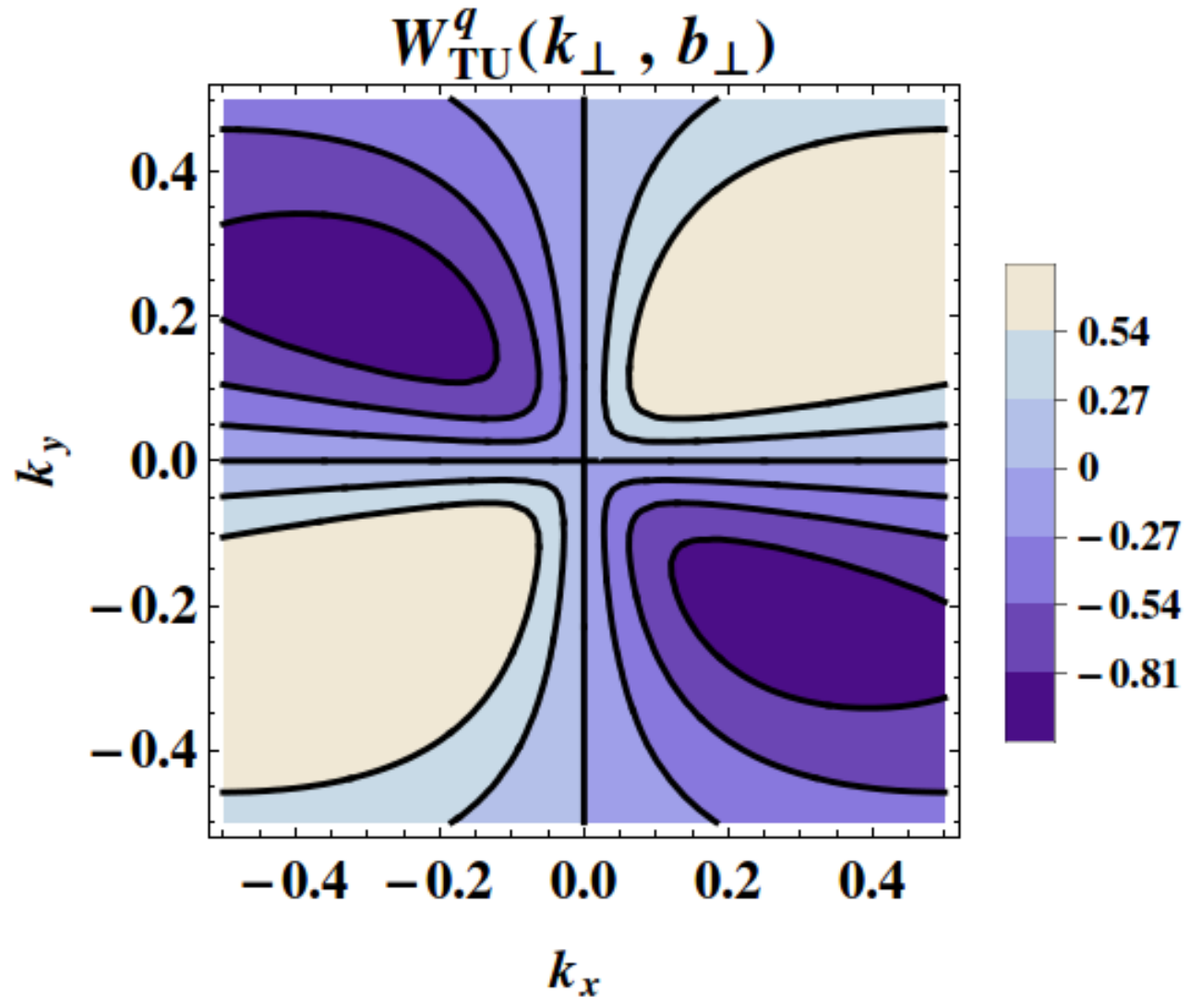}  
  \caption{Plot of quark Wigner distributions $W^q_{UU} ({\bs k_\perp}, {\bs b _\perp})$, $W^q_{LU} ({\bs k _\perp}, {\bs b _\perp})$ and $W^q_{TU} ({\bs k _\perp}, {\bs b _\perp})$ with $\Delta_{\perp max} = 20~GeV$. The left panel shows plot of these distributions in $b-$space with ${\bs k}_\perp= 0.4~\mathrm{GeV} \,\hat{{\bs e}}_y$  and  the right panel shows plot of these distributions in
  $k-$space with ${\bs b}_\perp= 0.4~ \mathrm{GeV}^{-1}$ \,$\hat{{\bs e}}_y$.}
  \label{quark}
\end{figure}
The Fourier transform is performed numerically using the Levin method \cite{Levin82} for integration which is well-suited for our oscillatory integrals, it gives converging results with respect to the change of the upper 
limit of the $\Delta_\perp = \Delta_{max}$ in the numerical integration. This is an improvement compared to our earlier works \cite{Mukherjee14,Mukherjee15}.
Figure.~\ref{quark} shows the contour plots for the quark Wigner distributions.
Fig.~\ref{quark} (a)--(c) are plots in impact parameter space for ${\bs k}_\perp= 0.4~\mathrm{GeV} 
\,\hat{{\bs e}}_y$.
Fig.~\ref{quark} (d)--(f) are plots in transverse momentum space for ${\bs b}_\perp= 0.4~\mathrm{GeV}^{-1} \,\hat{{\bs e}}_y$. One can also study Wigner distribtution in mixed space by integrating out $b_y$ and $k_x$ and by plotting the variables $b_x$ and $k_y$. But here we limit our discussion and plot the Wigner distributions only in transverse momentum and position space.
\begin{figure}[h]
 \centering 
(a)\includegraphics[width=5cm,height=3.5cm]{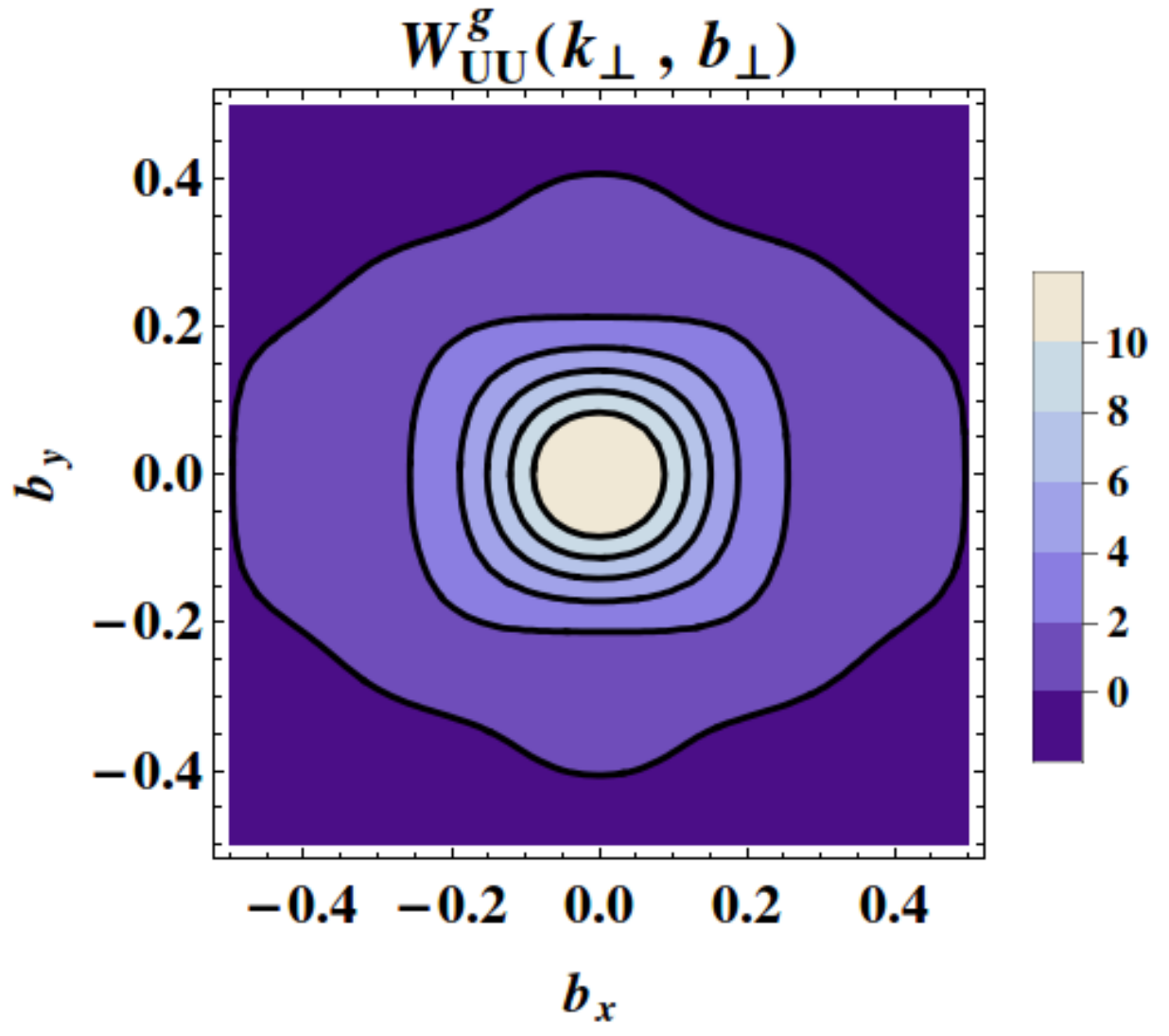} 
(d)\includegraphics[width=5cm,height=3.5cm]{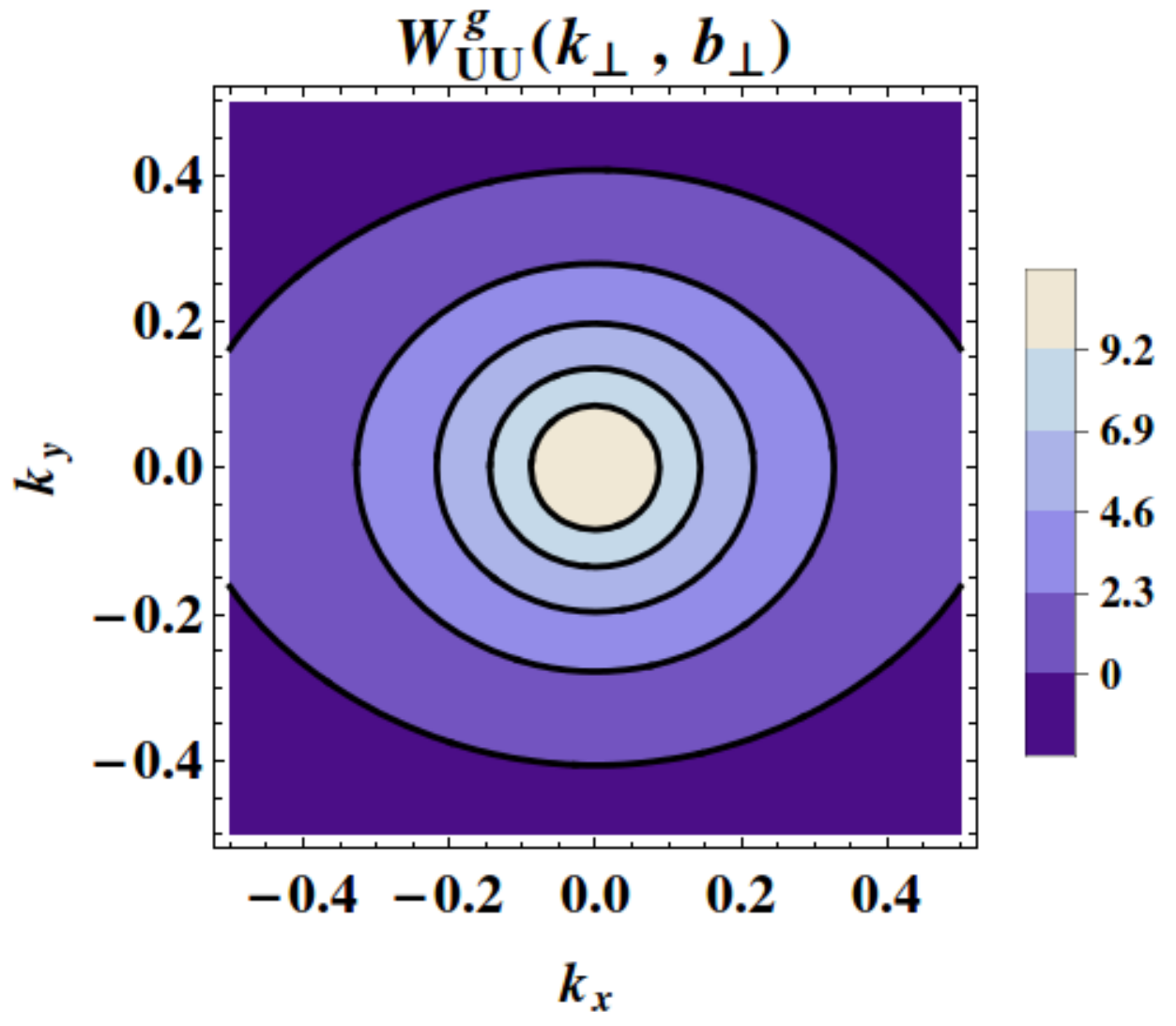}\\[1ex]
(b)\includegraphics[width=5cm,height=3.5cm]{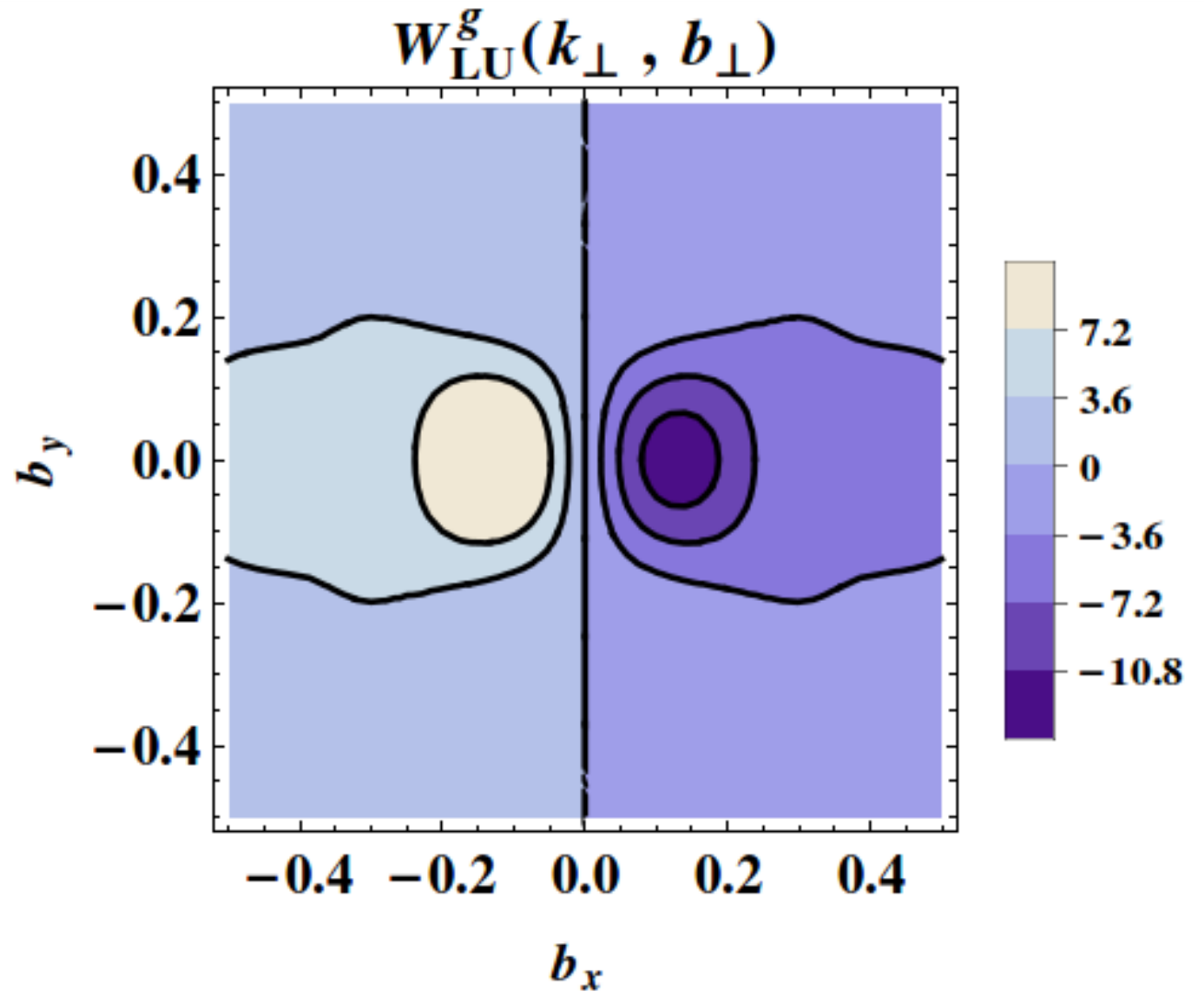}
(e)\includegraphics[width=5cm,height=3.5cm]{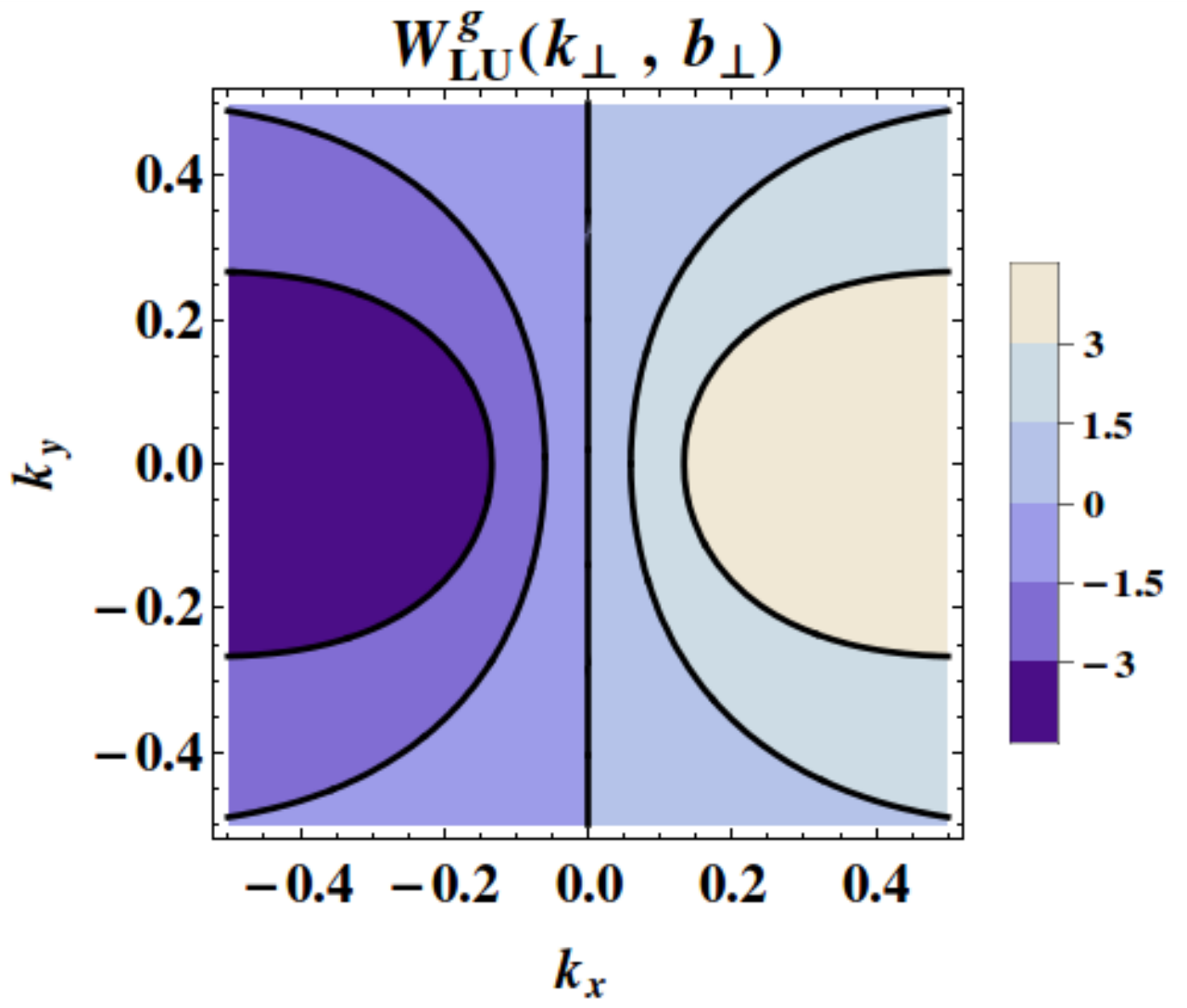}\\[1ex]
(c)\includegraphics[width=5cm,height=3.5cm]{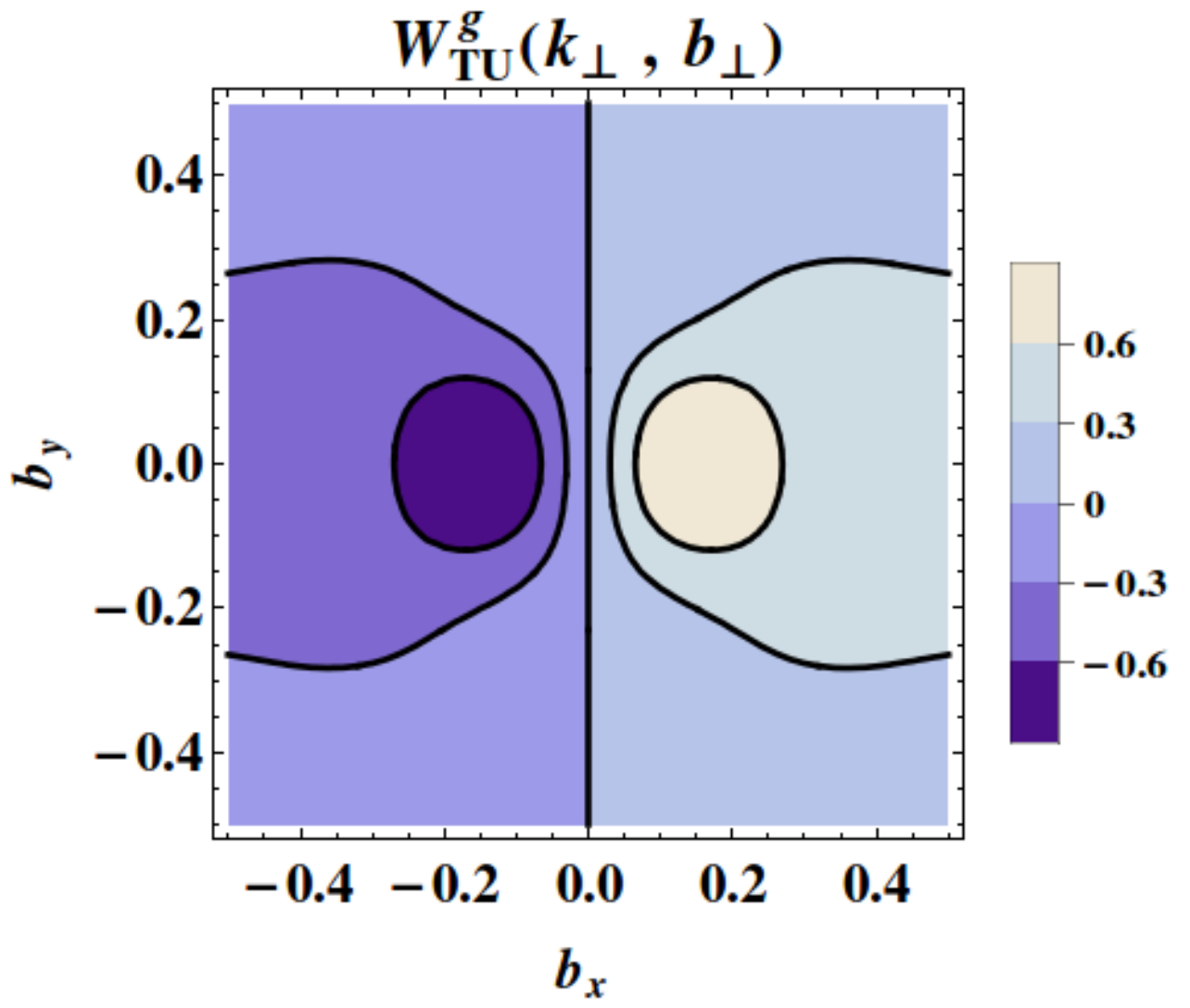}
(f)\includegraphics[width=5cm,height=3.5cm]{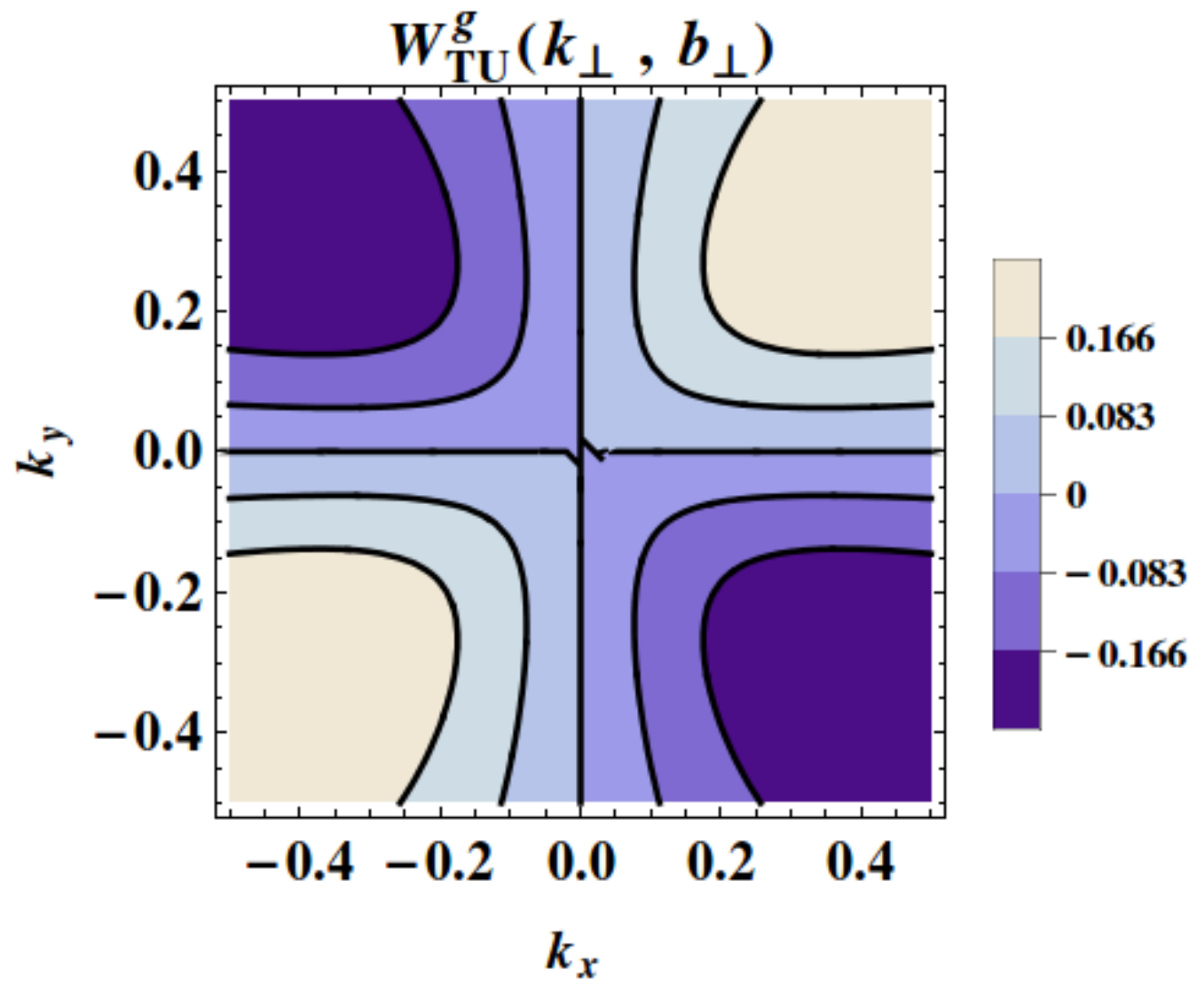}  
\caption{Plot of gluon Wigner distributions $W^g_{UU} ({\bs k _\perp}, {\bs b _\perp})$, $W^g_{LU} ({\bs k _\perp}, {\bs b _\perp})$ and $W^g_{TU} ({\bs k _\perp}, {\bs b _\perp})$ with $\Delta_{\perp max} = 20~GeV$. The left panel shows plot of these distributions in $b-$space with ${\bs k}_\perp= 0.4~\mathrm{GeV} \,\hat{{\bs e}}_y$  and  the right panel shows plot of these distributions in   $k-$space with ${\bs b}_\perp= 0.4~ \mathrm{GeV}^{-1}$ \,$\hat{{\bs e}}_y$.}
  \label{gluon}
\end{figure}
Fig.~\ref{quark} (a) shows the contour plot in ${\bs b _\perp}$ space for $W_{UU}^q({\bs k _\perp}, {\bs b _\perp})$ that represents the Wigner distribution for unpolarized quark in an unpolarized target state.
We see a positive peak centered in the ${\bs b _\perp}$ space. 
Fig.~\ref{quark} (d) gives the plot of  $W_{UU}^q({\bs k _\perp},{\bs b _\perp})$ in  ${\bs k_\perp}$ space. It shows a negative peak in the central region contrary to the ${\bs b _\perp}$ space.  
Fig.~\ref{quark} (b) shows the contour plot in ${\bs b _\perp}$ space for $W_{LU}^q({\bs k _\perp}, {\bs b _\perp})$ that represents the Wigner distribution for unpolarized quark in a longitudinally polarized target state. We observe a dipole-like nature in $W_{LU}^q({\bs k _\perp},{\bs b _\perp})$.
Similar dipole-like structure is observed for $W_{LU}^q({\bs k _\perp},{\bs b _\perp})$ in ${\bs k_\perp}$ space which is shown in Fig.~\ref{quark} (e).
Fig.~\ref{quark} (c) shows the Wigner distribution for unpolarized quark in a transversely polarized target state where the contour plot is shown for fixed value of ${\bs k _\perp}$ in
${\bs b _\perp}$ space. We again observe a dipole-like structure similar to $W_{LU}^q({\bs k _\perp},{\bs b _\perp})$.
In ${\bs k _\perp}$ space we observe a quadrupole-like structure.

Figure.~\ref{gluon} shows the contour plots for the gluon Wigner distribution.
Fig.~\ref{gluon} (a)--(c) are plotted in ${\bs b _\perp}$ space for fixed value of ${\bs k _\perp}$.
Fig.~\ref{gluon} (d)--(f) are plotted in ${\bs k _\perp}$ space  for fixed value of ${\bs b _\perp}$.
Fig.~\ref{gluon} (a) is similar in nature to the quark case since it is positive in the central 
${\bs b _\perp}$ space. But Fig.~\ref{gluon} (d) is positive whereas in the quark case we get a negative peak 
 in ${\bs k _\perp}$ space. Fig.~\ref{gluon} (b)and (e) which represent the case when the gluon is unpolarized and the target is longitudinally polarized are both showing a dipole-like structure similar to that observed in the quark case in ${\bs b}_\perp$ space and ${\bs k}_\perp$ space respectively. The nature of  $W_{TU}^g(x, {\bs k _\perp}, {\bs b _\perp})$ is also similar to that of the quark case. We see a dipole-like and quadrupole-like nature for the distributions $W_{TU}^g({\bs k _\perp}, {\bs b _\perp})$ in ${\bs b _\perp}$ and ${\bs k _\perp}$ space respectively.

\section{Conclusion}\label{conclusion}
We investigated quark and gluon Wigner distributions in a dressed quark model using an improved numerical technique that removes the cutoff dependence of $\Delta_\perp$ integration. Here we have discussed only the unpolarized quarks and gluons for different polarization of the dressed quark state at leading twist. The detailed version of quark and gluon Wigner distributions can be found in Refs.~\cite{More17,More18}. All the Wigner distributions can be expressed in terms of overlap of LFWFs and are then calculated using the analytic form of the LFWFs.
 We analyze them in terms of contour plots in the transverse position and momentum space.
\begin{acknowledgements}
We would like to thank Light Cone organizers for the invitation. AM and JM would like to thank the Department of Physics, IIT Bombay for providing financial support.
\end{acknowledgements}
%

\end{document}